\def\apj{Astroph. J.}
\def\aj{Astron. J.}
\def\apjl{Astroph. J. Lett.}\def\apjs{Astroph. J. Supp.}
\def\mnras{Mon. Not. R. Astron. Soc.}
\def\aap{Astron. Astroph. }

\def\nat{Nature}
\def\kms{km s$^{-1}$\/}
\documentclass{PoS}

\title{Narrow-line Seyfert 1s: what is wrong in a name?}

\ShortTitle{Narrow-line Seyfert 1s: what is wrong in a name?}

\author{\speaker{P. Marziani$^{1}$},
A. del Olmo$^{2}$, M. D' Onofrio$^{3}$, D.Dultzin$^{4}$, C. A. Negrete$^{4}$, M. L. Mart\'{\i}nez-Aldama$^{2}$, E.Bon$^{5}$, N. Bon$^{5}$, G. M. Stirpe$^{6}$\\
        $^{1}$INAF, Osservatorio Astronomico di Padova, Padova, Italy; $^{2}$ IAA-CSIC, Granada, Spain; $^{3}$Dipartimento di Fisica \& Astronomia, Universit\`a di Padova, Padova, Italy;  $^{4}$IA-UNAM, Mexico; 
        $^{5}$Belgrade Observatory, Serbia; $^{6}$ INAF, OASSB, Bologna, Italy \\
        E-mail: \email{paola.marziani@inaf.it}}

\abstract{Narrow-line Seyfert 1s (NLSy1s) are an ill-defined class. Work done over the past 20 years as well as recent analyses show a continuity in properties (e.g., Balmer line profiles, blueshifts of high-ionization lines) between sources with FWHM above and below 2000 \kms, the defining boundary of NLSy1s. This finding alone suggests that comparisons between samples of NLSy1s and rest of broad-line AGNs are most likely biased. NLSy1s can be properly contextualized by their location on the quasar main sequence originally defined by Sulentic et al \cite{sulenticetal00a}.  At one end, NLSy1s encompass sources with strong FeII emission and associated with high Eddington ratio that hold the promise of becoming useful distance indicators; at the other end, at least some of them are sources with broad profiles seen face-on. Any rigid FWHM limit gives rise to some physical ambiguity, as the FWHM of low-ionization lines depends in a complex way on mass, Eddington ratio, orientation, and luminosity. In addition, if the scaling derived from luminosity and virial dynamics applies to the broad line regions, NLSy1s at luminosity higher than 10$^{47}$ erg s$^{-1}$\ become physically impossible. Therefore, in a broader context, a proper subdivision of two distinct classes of AGNs and quasars may be achieved by the distinction between Pop. A and B with boundary at $\approx$ 4000 \kms in samples at $z <1$, or on the basis of spectrophotometric properties which may ultimately be related to differences in accretion modes if high-luminosity quasars are considered.}

\FullConference{Revisiting narrow-line Seyfert 1 galaxies and their place in the Universe - NLS1 Padova\\
		9-13 April 2018 \\
		Padova Botanical Garden, Italy}

\begin{document}

\section{Introduction}

The question of the title is, of course, rhetorical. There  is nothing wrong: it is perfectly legitimate to restrict to type-1 AGN with FWHM  H$\beta$\ $\le   2000$ \kms. Several slightly different definitions were proposed since the original paper of \cite{osterbrockpogge85}. We will adopt here the definition based on the FWHM limit, without consideration of FeII and OIII properties.  However, the limit  at FWHM(H$\beta$) $=$  2000 \kms  immediately raises several questions:  1) does this limit have  a well-defined  observational meaning? Or, in other words, is there any discontinuity between NLSy1s and  broad-line Sy1s (BLSy1s i.e., the rest of type-1 AGN with  FWHM(H$\beta$) $>$  2000 \kms) ? 2) Are NLSy1s a homogeneous class? 3) Can they be present in high-luminosity samples? 4) What is the physical meaning of a fixed FWHM(H$\beta$) limit?

We might ask whether the distinction between Population A (FWHM(H$\beta$) $\le$  4000 \kms at $\log L \lesssim 10^{46.5}$ erg s$^{-1}$\ which encompasses wind-affected \cite{richardsetal11} sources of moderate-to-high $L/L_\mathrm{Edd}$) and B \cite{sulenticetal00a} is more meaningful. In the following we will test the possibility that NLSy1s are a part of Population A, and show that this is indeed the case. We will compare NLSy1s vs. the rest of Population A quasars (hereafter RPopA), i.e., sources with $2000 < $ FWHM(H$\beta$) $\le 4000$ \kms.

 \section{NLSy1s as part of Population A}

NLSy1s  are part of  the type-1 quasar ``Eigenvector 1''   Main Sequence (MS), and are the ``drivers'' of the E1 correlations in low-$z$ samples since they have the narrowest H$\beta$\ line along the sequence, and,  are most frequent among the strong FeII emitters. If we subdivide the MS in spectral type in bins of FeII prominence (measured by the $R_\mathrm{FeII}$\ parameter, the ratio between the flux of the FeII blend at $\lambda$4570 \AA\ and  H$\beta$) as suggested by \cite{sulenticetal02}, NLSy1s span a broad range of spectral types, the $R_\mathrm{FeII}$ parameter being between almost 0 and 2 in the quasi-totality of sources. NLSy1s seamlessly occupy the low-end of the distribution of FWHM(H$\beta$) in the samples of \cite{shenetal11,zamfiretal10}. The $R_\mathrm{FeII}$ occupation of NLSy1s and RPopAs  is in the same range, although a larger fraction of strong FeII emitters is found among NLSy1s.  The ratio of the number of sources with $R_\mathrm{FeII}$\ in the range 1-2 with respect to the one with $R_\mathrm{FeII}$ in the range 0-1 is consistently higher in three different samples \cite{marzianietal03a,grupeetal99,shenetal11}, being 0.90 and 0.33 for NLSy1s and RPopA in the SDSS DR7 sample of \cite{shenetal11}, 0.21 and 0.04 in the type-1 AGN sample of \cite{marzianietal03a}, and 8.0 and 1.13 in a sample of soft-X ray bright  sources \cite{grupeetal99}. The last result may appear less surprising if one considers that \cite{grupeetal99} sources were selected at one extreme of the quasar MS  where high value of the photon index $\Gamma_\mathrm{S}$\ are frequently found, and that they are often associated with the most extreme FeII emitters. NLSy1s and RPopA show no apparent discontinuity as far as line profiles are concerned \cite{craccoetal16}.  Composite H$\beta$\ profiles of spectral types along the MS are consistent with a Lorentzian as do the RPopA sources. A recent  analysis  of composite line profiles in narrow FWHM(H$\beta$) ranges ($\Delta $ FWHM = 1000 \kms) confirms that there is no discontinuity at 2000 \kms; best fits of  profiles remain Lorentzian up to at least 4000 \kms. A change in shape occurs around FWHM(H$\beta$)  = 4000 \kms  at the Population A limit, not at the one of NLSy1s \cite{sulenticetal02}. No significant difference between CIV$\lambda$1549 centroids  at half-maximum of NLSy1s and rest of Pop. A is detected using the HST/FOS data of the \cite{sulenticetal07} sample.  

\section{Dis-homogeneity of NLSy1s}

The parameter $R_\mathrm{FeII}$ has emerged as an important Eddington ratio correlate \cite{grupeetal99,marzianietal01,duetal16a}. CIV blueshift amplitudes among NLSy1s depend on the spectral type along the MS, i.e., they increase with $R_\mathrm{FeII}$. If $R_\mathrm{FeII}$\ is larger than 1, the average blueshift  in the objects of \cite{sulenticetal07} exceeds -1000 \kms, but is almost 0 if $R_\mathrm{FeII}$ $\lesssim$0.5. Eddington ratio is correlated with $R_\mathrm{FeII}$\ \cite{duetal16a} in the sense that higher $R_\mathrm{FeII}$\ corresponds to higher  $L/L_\mathrm{Edd}$\ and spans a broad range in NLSy1s and RPopA as well ($0.1 - 0.2 \lesssim L/L_\mathrm{Edd}   \lesssim$ 1). 

\section{Luminosity effects}

Extreme Pop. A  (xA; RFeII$>$1) show  a broad range of FWHM beyond 2000 \kms
for virialized lines such as H$\beta$. In other words, there are high FWHM and high luminosity equivalents of NLSy1 with strong FeII emission. In addition, there is a minimum FWHM(H$\beta$) that is consistent with the virial assumption. The minimum FWHM(H$\beta$) increases with luminosity. Assuming a  virial relation $M_\mathrm{BH}  = f r_\mathrm{BLR} \mathrm{FWHM}^{2}/G$, where  BLR radius  $r_\mathrm{BLR} \propto  L^\mathrm{a}$, with $a \approx$  0.5-0.7, the minimum FWHM(H$\beta$) is well defined if there is a limiting Eddington ratio ($\sim 1$; \cite{marzianietal09}). Higher luminosity implies a displacement of the MS toward larger FWHM(H$\beta$) i.e., larger masses at fixed Eddington ratio. For example, Fig. 9 of \cite{sulenticetal17} shows that all sources with $L \gtrsim 10^{47}$ erg s$^{-1}$\ exceed FWHM(H$\beta$)$\approx$ 2000 \kms.
 
 \subsection{Weak-lined quasars and NLSy1s}
 
It is interesting to note that 70\%-80\% of weak-lined quasars [WLQ, W(CIV$\lambda$1549) $\le$10 \AA, \cite{diamond-stanicetal09}]   belong to extreme Pop. A \cite{luoetal15,marzianietal16a}. WLQs and Pop. xA show continuity in CIV shifts and equivalent widths WLQs appear as  extremes of extreme Population A. At low-$L$\ this means FeII-strong NLSy1s. However, since WLQs are high-luminosity sources, they show larger FWHM than NLSy1s; in the sample considered by \cite{marzianietal16a} only 1 source fully meets the defining criterion of  NLSy1s.

\begin{figure}[ht!]
\begin{center}
\includegraphics[width=15.5cm]{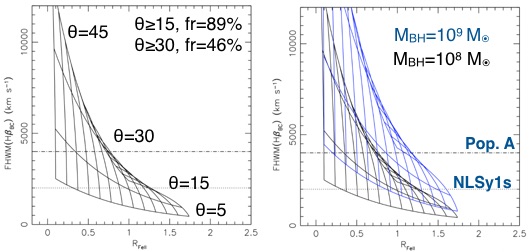}
\end{center}
\caption{The optical plane of the Eigenvector 1, FWHM(H$\beta$) vs. $R_\mathrm{FeII}$, with grids representing the source locations for $M_\mathrm{BH}=10^{8}$ M$_{\odot}$\ (left) and  for  $M_\mathrm{BH}=10^{8}$ M$_{\odot}$\ and $M_\mathrm{BH}=10^{9}$ M$_{\odot}$, respectively. Eddington ratio increases with increasing $R_\mathrm{FeII}$ from 0 to a few (in linear scale). In the left panel, the two grids are meant to show the effect of increasing the mass from  $M_\mathrm{BH}=10^{8}$ M$_{\odot}$\ to  $M_\mathrm{BH}=10^{9}$ M$_{\odot}$\ (blue).}
\label{fig:e1int} 
\end{figure}

\section{On the physical meaning of a fixed FWHM limit in the 4DE1 context}

The interpretation of the optical MS plane at low-$z$ is still debated \cite{marzianietal01,shenho14,pandaetal17,pandaetal18a}.  Eddington ratio and viewing angle may be the main factor if samples do not cover very large range of luminosity and redshift \cite{marzianietal01}. This suggestion has been recently repeated  \cite{shenho14} by an examination of the MS trends in a large SDSS sample. We propose here a toy scheme that explains in a qualitative way the occupation of the MS plane under the assumptions that Eddington ratio, mass and an aspect angle $\theta$ (i.e., the angle between the line-of-sight and the accretion disk axis)\ are the parameters setting the location of quasar along the MS. Under the standard virial assumption, we expect that FWHM(H$\beta$) $\propto M_\mathrm{BH}^{\frac{1}{4}} L/L_\mathrm{Edd}^{-\frac{1}{4}} f(\theta)^{-\frac{1}{2}}$, $R_\mathrm{FeII}$ $\propto \mathrm{R_\mathrm{FeII}} L/L_\mathrm{Edd} \cos \theta / (1 + b \cos\theta)$, where $R_\mathrm{FeII}$($L/L_\mathrm{Edd}$) is a relation that needs to be established either empirically or theoretically between $R_\mathrm{FeII}$ and Eddington ratio, $f$ is the form factor and $b$\ a limb darkening constant. For illustrative purposes, we considered the relations derived by \cite{duetal16a,rakshitetal17}. The relation between $R_\mathrm{FeII}$ and $L/L_\mathrm{Edd}$\ is especially uncertain at low $R_\mathrm{FeII}$ (and $L/L_\mathrm{Edd}$)\ because low $R_\mathrm{FeII}$ values are difficult to measure with good precision but also because the \cite{duetal16a} relation has been tested on sources that are FeII-strong. In addition, we ignore the fact that there are systematic differences in chemical composition along the MS, with the FeII-strong emitter  indicating highly-solar abundances \cite{negreteetal12}.  Figure \ref{fig:e1int} shows grids of lines as a function of Eddington ratio and orientation assuming the $R_\mathrm{FeII}$ - $L/L_\mathrm{Edd}$\ relation of \cite{duetal16a}. As expected, $\theta$\ and $L/L_\mathrm{Edd}$\ predominantly (but not exclusively) affect FWHM and $R_\mathrm{FeII}$, respectively.  In this intriguing that, under the assumptions of the toy scheme, and also following \cite{marzianietal01} the FWHM limit at 4000 \kms should include mainly sources  with $L/L_\mathrm{Edd} \gtrsim 0.1 - 0.2$, but also   radiators at lower Eddington ratio. However, they are expected to be rare because they should be observed almost pole-on ($P(\theta) \propto \sin \theta$). NLSy1s preferentially sample face-on sources along the MS. 

\section{Conclusion}

NLSy1s do not show a clear discontinuity with broader sources (BLSy1s) at least up to FWHM(H$\beta$)  = 4000 \kms\ in low-$z$ samples. It is apparently more meaningful to talk of Population A and B since the distinction is more closely associated to a critical $L/L_\mathrm{Edd}$. The caveats  previously analyzed imply that only few pole-on Population B sources should ``enter'' into the FWHM domain of Pop. A, at low $R_\mathrm{FeII}$.  

Meaning of inter-comparison  NLSy1s -- BLSy1s is difficult to assess.  Increasing luminosity implies an  increase in the minimum FWHM(H$\beta$); at $\log L \gtrsim 47$ [erg s$^{-1}$] NLSy1s are not anymore possible. Sources with $R_\mathrm{FeII}$ $>1$\ (xA) show fairly constant properties, with small dispersion around a well-defined Eddington ratio, regardless of their line width. WLQs can be explained as extreme xA sources.

Assuming a fixed limit in FWHM exposes samples to various biases.  Pop. A preferentially  isolates higher Eddington ratio sources. NLSy1s are not a uniform class in terms of physical properties but they may properly considered as the face-on sources along the MS.  

Are NLSy1s quasars in their early stage of the evolution, akin to high-$z$ quasars \cite{sulenticetal00a,mathur00}?  The ontogeny of black holes is represented by their monotonic increase in mass. Their phylogeny involves  Population A (and NLSy1s as a part of Population A): Population A may be seen as a progenitor to  more massive (evolved) radio-quiet Population B sources \cite{fraix-burnetetal17}. The same considerations could be applied also to the rare RL NLSy1 \cite{komossaetal06} that is, they may evolve into massive, powerful RL quasars  \cite{lister18}.


\section*{Acknowledgements}

A.d.O. and M.L.M.A. acknowledge financial support from the Spanish AEI and  European FEDER fundings through the research project AYA2016-76682-C3-1-P. D.D. and A.N. acknowledge support from grants PAPIIT108716, UNAM, and CONACyT221398, P.M. and  M. D. O.  from  the INAF PRIN-SKA 2017, program 1.05.01.88.04. This conference has been organized with the support of the
Department of Physics and Astronomy ``Galileo Galilei'', the 
University of Padova, the National Institute of Astrophysics 
INAF, the Padova Planetarium, and the RadioNet consortium. 
RadioNet has received funding from the European Union's
Horizon 2020 research and innovation programme under 
grant agreement No~730562. 

\bibliographystyle{JHEP}
\providecommand{\href}[2]{#2}\begingroup\raggedright\endgroup


\end{document}